\documentclass[a4paper]{article}
\usepackage{amsmath,amssymb,xcolor,tikz-cd}
\usepackage[affil-it]{authblk}
\usepackage[smalltableaux]{ytableau}
\usepackage[final, colorlinks=true,linkcolor=blue!80,citecolor=blue!80]{hyperref}
\usepackage{cleveref,orcidlink}
\usepackage[final,spacing]{microtype}
\usepackage[textsize=tiny]{todonotes} 
\usepackage[osf]{newpxtext}
\usepackage[varbb,osf,slantedGreek]{newpx}
\usepackage[russian,british]{babel}
\newcommand{\lb}[1]{\todo[color=green!20]{LB: #1}}

\newcommand\email[1]{\href{mailto:#1}{\textcolor{blue!80}{\textup{\texttt{#1}}}}}
\renewcommand\Omega\Omegaup
\renewcommand\delta\deltaup

\begin{document}
\title{Gravity from AKSZ--Manin theories in two, three, and four dimensions}
\author[1,2]{Leron Borsten\orcidlink{0000-0001-9008-7725}}
\author{Dimitri Kanakaris\orcidlink{0009-0001-7716-851X}}
\author{Hyungrok Kim\orcidlink{0000-0001-7909-4510}}

\affil{Department of Physics, Astronomy and Mathematics, University of Hertfordshire, Hatfield, Herts.\ \textsc{al10 9ab}, United Kingdom}
\affil[2]{Blackett Laboratory, Imperial College London
Prince Consort Road, London \textsc{sw7 2az}, United Kingdom}
{
    \makeatletter
    \renewcommand\AB@affilsepx{: \protect\Affilfont}
    \makeatother

    \makeatletter
    \renewcommand\AB@affilsepx{, \protect\Affilfont}
    \makeatother

\affil{\email{l.borsten@herts.ac.uk}}
    \affil{\email{d.kanakaris-decavel@herts.ac.uk}}
    \affil{\email{h.kim2@herts.ac.uk}}
}

\maketitle

\begin{abstract}
We show that various \emph{dynamical} gauge theories in two, three and four dimensions,  obtained as \emph{Manin} deformations of  topological Alexandrov--Kontsevich--Schwarz--Zaboronsky  (AKSZ) theories, are equivalent to gravitational theories. Since  gravity is topological in two and three dimension, this equivalence requires a diffeomorphism breaking background source term. In four dimensions,  however, the MacDowell--Mansouri formulation of Einstein  gravity with a cosmological constant is obtained identically via the Manin deformation of a particular AKSZ theory. 
\end{abstract}
\tableofcontents
\section{Introduction and summary}

In \cite{Borsten:2024pfz} it was shown that three-dimensional Einstein gravity with a cosmological constant and a (diffeomorphism-breaking) background stress--energy tensor density is equivalent to  a Manin theory. The latter, in turn,  is essentially a three-dimensional Chern--Simons theory with a certain mass-like deformation, which renders the theory non-topological and \emph{dynamical}. In the simplest case, the Manin theory  is  equivalent to $\textrm{SL}(2;\mathbb R)$ Yang--Mills theory and we arrive at the surprising conclusion that, with a diffeomorphism-breaking background stress--energy tensor density, gravity \emph{is} Yang--Mills. This equivalence is at the level of the actions, which are identical. However, one may still distinguish them by insisting on, or relaxing, the invertibility of the connection associated to translations. 

In turn, Chern--Simons theory is the special case of the class of Alexandrov--Kontsevich--Schwarz--Zaboronsky (AKSZ) topological field theories\footnote{See e.g.~ \cite{Roytenberg:2006qz,Cattaneo:2010re,Kotov:2010wr,Ikeda:2012pv,Kim:2018wvi} for reviews of AKSZ theories.} associated to an (ungraded) Lie algebra \cite{Alexandrov:1995kv}. Thus,  replacing the strict Lie algebra with more general \(L_\infty\)-algebras or \(L_\infty\)-algebroids suggests  natural  higher gauge theory\footnote{See e.g.~\cite{Jurco:2018sby, Borsten:2024gox} and references therein for reviews of higher gauge theory.} analogs of the construction given in \cite{Borsten:2024pfz}. These \emph{higher AKSZ--Manin theories} ought to  correspond to gauge/gravity theories in other dimensions. Indeed,  here we show that, in two, three, and four spacetime dimensions, for judicious  choices of gauge algebras, they are equivalent to theories of gravity coupled to backgrounds in various ways:
\begin{enumerate}
\item Two dimensions: Jackiw--Teitelboim (JT) gravity with background stress--energy tensors or  torsion.
\item Three dimensions: (Anti)-de Sitter Einstein gravity and Einstein--Cartan-like gravity with background stress-energy tensors.  
\item Four dimensions: the $BF$ formulation of MacDowell--Mansouri gravity with or \emph{without} a background stress--energy tensor.
\end{enumerate}
The general structure in all cases is that of an AKSZ theory with graded tangent Lie algebra $\mathrm T^*[d-3]\mathfrak{iso}(d)$, where $\mathfrak{iso}(d)$ is the local isometry algebra of the $d$-dimensional spacetime (with Euclidean or Lorentzian signature), i.e.~the Poincar\'e, de Sitter (dS) or anti-de Sitter (AdS) algebras  according as the cosmological constant $\Lambda$ is vanishing, positive or negative.  

However, although uniformly constructed, four dimensions is subtly  different since gravity becomes dynamical.  In two and three dimensions, JT and Einstein-Cartan gravity are topological to begin with and there exist AKSZ formulations of these theories; the Manin deformations in these cases break their topological character and  induce dynamics (equivalent, for example, to a non-linear sigma model with AdS target or $\textrm{SL}(2;\mathbb R)$  Yang--Mills theory).  On the other hand, in four dimensions Einstein gravity is dynamical, whereas the undeformed AKSZ theory is topological. So, there can be  no direct AKSZ formulation of gravity;  the Manin deformation rectifies precisely this mismatch.  Indeed, the corresponding AKSZ--Manin theory  is precisely  the \(BF\) formulation \cite{Smolin:2003qu,Freidel:2005ak,Starodubtsev:2005mf,Freidel:2006hv}  of MacDowell--Mansouri gravity \cite{MacDowell:1977jt,Stelle:1979va} (reviewed in \cite{Wise:2006sm,Langenscheidt:2019qje}), which is classically equivalent to Einstein gravity with a cosmological constant. As in one and two dimensions,   it is possible to further introduce a background stress-energy tensor through the Manin deformation. 

As mentioned in \cite{Borsten:2024pfz}, this equivalence holds at the classical, perturbative level. For the nonperturbative quantum theory, one needs to deal with issues such as  degenerate metrics, summing over topologies, and large gauge transformations, which we do not discuss in this paper.

Note, there is significant  prior work on deformed-AKSZ realisations of non-topological  theories, see for example \cite{Barnich:2010sw, Grigoriev:2010ic, Grigoriev:2012xg, Alkalaev:2013hta, Grigoriev:2016wmk, Pulmann:2019vrw, Grigoriev:2020xec}.  In particular, MacDowell-Mansouri--Stelle--West type formulations of gravity \cite{MacDowell:1977jt,Stelle:1979va} in four and higher dimensions have been derived from a presymplectic\footnote{Allowing degenerate symplectic form.} AKSZ  perspective in \cite{Alkalaev:2013hta}.  Although the  gauge structures and corresponding fields of the  deformed  AKSZ constructions of MacDowell-Mansouri gravity given in \cref{sec:4d} here and in \cite[\S 3.5.1]{Alkalaev:2013hta} are different\footnote{Specifically,    here we take $\mathfrak{so}(2, d-1)[d-1]\to\mathfrak{so}(2, d-1)$ as the gauge algebra, while in \cite{Alkalaev:2013hta} the initial gauge algebra is $\mathbb{R}^{d+1}\to\mathfrak{so}(d-1,2)[1]\to \mathfrak{so}(d-1,2)[2]$, which is then reduced to  $\text{AdS}_{d+1}\to \mathfrak{so}(d-1,2)[1]\to \mathfrak{so}(d-1,2)[1]$ (note, we have switched to our grading conventions for direct comparison). Correspondingly, here we have unconstrained $\mathfrak{so}(2, d-1)$-valued $1$- and $2$-forms, while in \cite{Alkalaev:2013hta} there are    $\mathfrak{so}(2, d-1)$-valued $1$- and $2$-forms and an  $\mathbb{R}^{d+1}$-valued $0$-form, where the  $2$-form and $0$-form are constrained to effect the embedding $\text{AdS}_{d+1} \hookrightarrow \mathbb{R}^{d+1}$ and to ensure the connection is torsion free.}, both capture precisely the same physics and it would be interesting to understand how they are related. Since  the respective approaches to deforming the underlying AKSZ formalism differ, this connection is not immediately obvious, but one possible approach would be  to pass through an $L_\infty$-span as in \cite{JalaliFarahani:2023sfq}. Importantly, it should also be noted that by allowing a countable infinite tower of auxiliary fields, it is possible to construct a genuine AKSZ formulation for \emph{any} gauge theory, including gravity as developed in \cite{Barnich:2010sw, Grigoriev:2010ic, Grigoriev:2012xg}. 

Another approach  for describing theories with degrees of freedom in an AKSZ-like manner is given in \cite{Bonechi:2022aji},  building on the  equivariant BV framework of \cite{Bonechi:2019dqk} and  a generalized
AKSZ approach to abelian Yang–Mills theory developed in \cite[chapter 5]{costello2011renormalization}. In \cite{Bonechi:2022aji} an AKSZ construction of a BV Donaldson--Witten theory is given. By means of the equivariant BV formalism and the associated BV push-forward, the BV Donaldson--Witten theory is shown to be formally quasi-isomorphic to an effective Yang-Mills action, providing a  consistent non-local deformation of   \cite[chapter 5]{costello2011renormalization} satisfying the  equivariant master equation. Consequently,  simple local operators in one theory can be mapped to  non-local ones in the other. 

 Finally, one can construct non-topological theories in $d$ dimensions as the boundary to a $(d+1)$-dimensional topological AKSZ `sandwich' theory as shown in \cite{Pulmann:2019vrw} using the notion of a relative field theory \cite{Freed:2012bs}. It would be interesting to connect the Manin deformations in $d$ dimensions to the non-topological boundary conditions in \cite{Pulmann:2019vrw}, which are  determined by  differential graded Lagrangian submanifolds of the AKSZ BV-manifold (as opposed to the Manin deformations used here, which are determined by Lagrangian submanifolds of the AKSZ target space). Establishing such a relation  would potentially allow for the formulation of \emph{dual} AKSZ--Manin theories, obtained via dual topological  conditions on the other boundary of the sandwich theory.

Looking further ahead, these constructions also suggest  holographic dualities in the presence of backgorunds. For instance, Jackiw--Teitelboim gravity with a negative cosmological constant is well known to admit an AdS\textsubscript2/CFT\textsubscript1 holographic interpretation \cite{Sachdev:2010um,2015escq.progE...2K} in terms of the Sachdev--Ye--Kitaev model \cite{Sachdev:1992fk,2015escq.progE...2K} and the Schwartzian action  (reviewed in \cite{Mertens:2022irh}). It will be interesting to see whether our constructions allow a holographical interpretation even when a background is present.

\paragraph{Conventions}
Our metric signature is always \(-+\dotsb+\), with \(\epsilon_{012\dotsb}=-\epsilon^{012\dotsb}=1\). Our \(L_\infty\)-algebras are such that the \(i\)-ary bracket \(\mu_i\) carries degree \(2-i\) and is totally antisymmetric.

\section{Review of Manin theories}

\subsection{Chern--Simons Manin theories in three dimensions}

Here we briefly review the Chern--Simons-type Manin theories introduced in \cite{Arvanitakis:2024dbu}. They are deformations of conventional  Chern--Simons theories with gauge Lie algebra $\mathfrak d$, where the deformation is by a mass-like term controlled by the data of a  \emph{Manin pair}  $(\mathfrak d,\mathfrak g)$ \cite{Dri89}. Notable examples of Manin theories include conventional Yang--Mills theory  and  Freedman--Townsend theory \cite{Freedman:1980us}. In both cases the deformation renders the initially topological theory dynamical, with \emph{massless} propagating degrees of freedom; the superficially mass-like deformation is in fact an algebraic quadratic term for auxiliary fields valued in a complement of $\mathfrak  g \subset\mathfrak d$.

\paragraph{Manin pairs and Hodge structures} A Manin pair  \((\mathfrak d,\mathfrak g)\) is a Lie algebra \(\mathfrak d\) equipped with an invariant nondegenerate inner product \(\langle-,-\rangle\) of split signature and a Lie subalgebra \(\mathfrak g\subset\mathfrak d\) that is maximal isotropic, i.e.\ whose dimension is half that of \(\mathfrak d\) and such that \(\langle x,y\rangle=0\) whenever \(x,y\in\mathfrak g\). 

A Hodge structure on a Manin pair is a linear map \(M\colon \mathfrak d\to\mathfrak d\)
 of mass dimension \(1\) with \(\operatorname{im}M\subset\mathfrak g\subset\ker M\) such that
\begin{equation}
\label{M axioms}
    \langle My,z\rangle=\langle y,Mz\rangle,\qquad 
    M[x,y]=[x,My]
\end{equation}
for \(x\in\mathfrak g\) and \(y,z\in\mathfrak d\). 
An immediate consequence of these axioms is that \(\langle-,M-\rangle\) is a (degenerate) symmetric bilinear form which is \emph{\(\mathfrak{g}\)-invariant},
\begin{equation}
\label{g-invariance}
	\langle[x,y],Mz\rangle + \langle y,M[x,z]\rangle = 0\,.
\end{equation}
Note that if the relation \(\mathfrak{g} \subset \ker M\) holds strictly, \(\mathfrak{g}\)-invariance may not be exhaustive.

\paragraph{Chern--Simons Manin theories} Consider a conventional (three-dimensional) Chern--Simons theory with gauge Lie algebra \(\mathfrak d\), whose action is then
\begin{equation}
	S_{\text{CS}}[\mathbb A] = \int\tfrac12k\left\langle\mathbb A,\mathrm d\mathbb A + \tfrac13[\mathbb A\wedge\mathbb A]\right\rangle, 
\end{equation}
where  \(\mathbb A\) is a \(\mathfrak d\)-valued Chern--Simons one-form.

The Chern--Simons Manin theories, as introduced in \cite{Arvanitakis:2024dbu},  are mass-like  Manin pair \((\mathfrak d,\mathfrak g)\) deformations of $S_{\text{CS}}[\mathbb A]$ given by
\begin{equation}
	S_{\text{CS}}^{\text{Manin}}[\mathbb A] = S_{\text{CS}}[\mathbb A]+\int\tfrac12\left\langle\mathbb A,\hat\star M\mathbb A\right\rangle.
\end{equation}
  Note that the mass-like term requires a Hodge star \(\hat\star\), which is taken with respect to a background (pseudo-)Riemannian metric \(\hat g\). Consequently, the  mass-like term breaks diffeomorphism symmetry down to the subgroup of isometries of \(\hat g\), and also breaks the gauge symmetries down to   \(\mathfrak g\subset\mathfrak d\) since $\ker(M)\cong\mathfrak g$.

To unpack this construction further, let   \(\tilde{\mathfrak g}\) be a linear subspace inside \(\mathfrak d\) such that \(\mathfrak d=\mathfrak g+\tilde{\mathfrak g}\).\footnote{We use $+$ for the direct sum of vector spaces, preserving $\oplus$ for the direct sum of Lie algebras.}
Then  \(\langle-,-\rangle\) identifies \(\tilde{\mathfrak g}\cong\mathfrak g^*\), and  we may  choose bases \(\{t_a\}\)  and   \(\{\tilde t^a\}\)  for \(\mathfrak g\)  and \(\tilde{\mathfrak g}\), respectively,  such that $\langle\tilde t^a,t_b\rangle = \delta^a_b$. In this case the structure constants of the Manin pair and the operator \(M\) are given by
\begin{equation}
\begin{aligned}
    [t_a,t_b]&=f_{ab}{}^ct_c,&
    [\tilde t^a,\tilde t^b]&=\tilde f^{ab}{}_{c}\tilde t^c+\tilde h^{abc}t_c,&
    [t_a,\tilde t^b]&=\tilde f^{bc}{}_at_c-f_{ac}{}^b\tilde t^c, 
\end{aligned}
\end{equation}
and\footnote{Note, $M\tilde t^a=M^{ab}t_b$  transforms in a   (sub)representation of that carried by $\tilde{\mathfrak{g}}$.} 
\begin{equation}
 Mt^a=0, \qquad  M\tilde t^a=M^{ab}t_b.
\end{equation}
The \(\mathfrak d\)-valued gauge field \(\mathbb A\) decomposes as \(\mathbb A=A^at_a+\tilde A_a\tilde t^a\) and we have 
\begin{multline}\label{eq:Chern--Simons-manin-action}
    S[A,\tilde A] =
    \int k\tilde A_a\wedge F^a + \tfrac12k\tilde f^{ab}{}_c\tilde A_a\wedge\tilde A_b\wedge A^c \\+ \tfrac16k\tilde h^{abc}\tilde A_a \wedge\tilde A_b \wedge\tilde A_c +\tfrac12M^{ab}\tilde A_a\wedge\hat\star\tilde A_b,
\end{multline}
where we have defined the field strength \(F = F^at_a = \mathrm dA + A\wedge A\) whose components are \(F^a=\mathrm dA^a+\frac12f_{bc}{}^aA^b\wedge A^c\). Choosing appropriate \((\mathfrak d,\mathfrak g)\) and $M$, ones recovers a variety of well-known gauge theories \cite{Arvanitakis:2024dbu}, which in special cases are Einstein gravity with a diffeomorphism-breaking background source \cite{Borsten:2024pfz}.

Let us illustrate  explicitly how the Hodge  structure \(M\) dictates the gauge structure of the theory. A gauge transformation of the action by a local \(\mathfrak{d}\)-valued gauge parameter \(\mathbb{c}\) gives
\begin{equation}
	\delta_{\mathbb{c}}S[\mathbb{A}]
	=
	\int \left\langle \mathbb{A},\hat\star M\delta_{\mathbb{c}}\mathbb{A}\right\rangle
	=
	\int \left\langle \mathbb{A},\hat\star M\big(-d\mathbb{c} + [\mathbb{c},\mathbb{A}]\big)\right\rangle.
\end{equation}
Assuming \(\ker M \cong \mathfrak{g} \subset \mathfrak{d}\),  implies  \(Md\mathbb{c} = 0\) iff \(\mathbb{c}\) is valued in \(\ker M\). Thus, we find that that the gauge algebra is a subalgebra of \(\ker M \cong \mathfrak{g}\). Imposing this is {necessary} for gauge invariance of the action. It is also sufficient; indeed, this follows from \(\mathfrak{g}\)-invariance \eqref{g-invariance} of the bilinear form \(\langle-,M-\rangle\):
\begin{align}
	&\left.
	\begin{aligned}
		\int\langle\mathbb{A},\hat\star Md\mathbb{c}\rangle &= 0
		&
		&\text{($\mathbb{c}$ $\mathfrak{g}$-valued)}
		\\
		\int\langle\mathbb{A},\hat\star M[\mathbb{c},\mathbb{A}]\rangle &= 0
		&
		&\text{($\langle-,M-\rangle$ $\mathfrak{g}$-invariant)}
	\end{aligned}
	\right\}
	&
	&\Leftrightarrow
	&
	\delta_{\mathbb{c}}S[\mathbb{A}] &= 0\,.
\end{align}
When we take the relation \(\mathfrak{g} \subset \ker M\) to hold \emph{strictly}, we find that the gauge group may in fact be larger. However, reading off the precise structure of this group would require additional information to be given about the mass operator \(M\).

\subsection{AKSZ--Manin theories in arbitrary dimensions}

Chern--Simons theory is a special instance of the AKSZ construction \cite{Alexandrov:1995kv}, which may be employed in an arbitrary number of spacetime dimensions. This facilitates the natural generalisation of the Chern--Simons Manin theories to any dimension as \emph{AKSZ--Manin theories}, the general theory of which is  developed in \cite{Arvanitakis:2025nyy}.

The AKSZ construction yields  solutions to the classical BV master equation directly from the data of source and target, $\Sigma$ and   $X$, which are NQ- and symplectic NQ-manifolds, respectively. The result is the classical BV action for a topological sigma model, where typically $\Sigma$ is the grade-shifted tangent bundle over  the world-volume or spacetime (depending on context) and $X$ is a (generalised)  target space. The BV fields (including   anti-fields, ghosts and anti-field ghosts) are then given by maps from $\Sigma$ to   $X$.  We shall not need the full machinery here, as we are merely concerned with the classical actions obtained from the AKSZ construction by setting all but the degree 1 fields (i.e.~the physical fields, which may be auxiliary, as opposed to the ghosts and antifields) to zero.

For our purposes, it is sufficient to regard the result as a theory of pairs of $p$- and $(d-p-1)$-form fields on a $d$-dimensional spacetime $\Sigma$. Since the physical $p$-form fields carry total degree $1$, they are  valued in the $(1-p)$-degree component $V^{(1-p)}$ of the (degree-wise finite) graded vector space $V=\bigoplus_{n} V^{(n)}$  of  an $L_\infty$-algebra\footnote{In the presence of scalar fields, this may generalised to an $L_\infty$-algebroid since they may be valued in a nontrivial manifold.} 
\begin{equation}
\mathfrak L = (V, \mu_k),
\end{equation}
where $\mu_k\colon V\times \cdots \times V \to V$ are the $k$-linear $L_\infty$-brackets. This $L_\infty$-algebra is the (higher) gauge algebra of the theory.

Letting $q=d-p-1$, each  $p$-form field valued in $V^{(1-p)}$ must be paired with a    $q$-form field valued in $V^{(1-q)}=V^{(2-d+p)}$. Thus, we require   $V^{(1-q)}\cong V^{(1-p)}{}^*$ to ensure the pairing exists (which follows from the fact that $\mathfrak L$ carries a cyclic structure of degree \(d-3\)).

Choosing a basis $\{t^{i_p}\}$, where $i_p=1, 2, \ldots \dim V^{(1-p)}$, for each $V^{(1-p)}$, the corresponding $p$-form field may be written $\phi_{(p)} = \phi_{i_p}t^{i_p}$. Let $t^{i_q} = t_{i_p}$, where  $\{t_{i_p}\}$ is the  basis of $V^{(1-p)}{}^*\cong V^{(1-q)}$ such that $\langle t_{i_p}, t^{j_p} \rangle=\delta_{i_p}{}^{j_p}$. Then the $q$-form field $\phi_{(q)}$ paired with $\phi_{(p)}$ may be written $\phi_{(q)} = \phi_{i_q}t^{i_q} = \phi^{i_p}t_{i_p} $.

With these conventions, the (classical part of) AKSZ action can  be written very schematically as 
\begin{equation}
S_{\text{AKSZ}}= \int \frac12\sum_{p=0}^{d-1}\phi_{i_p}\wedge d \phi^{i_{d-p-1}} +\sum_{p_1+\cdots+p_k=d} \frac{1}{k!}\mu^{i_{p_1}}{}_{i_{p_1}\cdots i_{p_k}} \phi_{i_p}\wedge  \phi^{i_{p_2}}\wedge \cdots \wedge \phi^{i_{p_k}},
\end{equation}
where  $\mu^{i_{p_1}}{}_{i_{p_1}\cdots i_{p_k}}$ are the structure constants of the $L_\infty$-brackets $\mu_k$ (and we have assumed $\mu_1 = \mathrm{id}$). 

To realise three-dimensional Chern--Simons theory with metric Lie algebra $\mathfrak{g}$ as an AKSZ theory, simply set $d=3$ and let $\mathfrak L= \mathfrak g$. In this case we only have one-forms $A_i$ and $\mu_2(-, -)=[-,-]$, so we recover the Chern--Simons action.

In direct analogy with the Chern--Simons case, to construct an AKSZ--Manin theory one needs an $L_\infty$-subalgebra $\mathfrak L'$ and a map $M\colon \mathfrak L\to \mathfrak L$ with kernel containing $\mathfrak L'$ and image contained in \(\mathfrak L'\).  Then we may deform the AKSZ action by a mass-like term 
\begin{equation}
S^{M}_{\text{AKSZ}}= S_{\text{AKSZ}} + \frac12\int \sum_{p=0}^{d-1}\phi_{i_p}\wedge  \hat \star M^{i_pj_p}\phi_{j_p}.
\end{equation}
For this deformation to be well-defined, these structures must  obey certain constraints as determined  in \cite{Arvanitakis:2025nyy}, which reduce to those of a Manin pair and Hodge structure in $d=3$.  Deriving the remaining gauge structure is directly analogous to the Chern-Simons case. Similarly, one of gauge algebra  \(\operatorname{im} M = \mathfrak{g} = \operatorname{ker}M\), which is  an \textit{admissible} \(L_\infty\)-subalgebra \(\mathfrak{g}\). In this case, much like the Chern-Simons  case, adding a Manin deformation  will  break the gauge algebra to the admissible \(L_\infty\)-subalgebra \(\mathfrak{g}\). Full details may be found in \cite{Arvanitakis:2025nyy}.

All examples we shall consider here automatically satisfy the required conditions, as evidenced by the consistency of the resulting gauge theories. 

\section{Two dimensions}\label{sec:2d}
We first consider theories in two dimensions, where the source  $\Sigma$ is taken to be a two-dimensional  (pseudo-)Riemannian spacetime manifold. Here  \(\mu,\nu,\dotsc\in\{0,1\}\) are  world indices, and \(a,b,\dotsc\in\{0,1\}\) are zweibein indices, while \(i,j,\dotsc\in\{0,1,2\}\) are adjoint indices of \(\mathfrak{sl}(2;\mathbb R)\).

We take as the $L_\infty$-algebra $\mathfrak{L}$ the graded Lie algebra
\begin{equation}
    \mathrm T[-1]^*\mathfrak{sl}(2;\mathbb R)
    =
    \mathfrak{sl}(2;\mathbb R)\ltimes\mathfrak{sl}(2;\mathbb R)^*[-1],
\end{equation}
which only has degree $0$ and $1$ components\footnote{Here we are using the degree-shift notation $(V[i])^{(k)}=V^{(i+k)}$, so that in particular if $V$ is ungraded  then  $V[i]$ only has degree $-i$.} and  commutators 
\begin{equation}
\left[t_i, t_j\right]=f_{ij}{ }^k t_k, \quad\left[t_i, \tilde{t}^j\right]=-f_{i k}{ }^j \tilde{t}^k, \quad\left[\tilde{t}^i, \tilde{t}^j\right]=0,
\end{equation}
where  \(\{t_i\}_{i=0}^{2}\)  and \(\{\tilde t^i\}_{i=0}^{2}\) are  some \(\mathfrak{sl}(2;\mathbb R)\) and \(\mathfrak{sl}(2;\mathbb R)^*\) bases, respectively. Note that the final commutator vanishes for degree reasons; also, in the gravitational interpretation given in \cref{ssec:2dgrav},  $\mathfrak{sl}(2;\mathbb R)\cong \mathfrak{so}(1,2)$ should be thought of as the (anti-)de Sitter algebra. 

On $ \mathrm T[1]^*\mathfrak{sl}(2;\mathbb R)$, we have an invariant inner product of degree \(-1\) given with respect to the above basis by  
\begin{equation}
\langle t_i, \tilde t^j \rangle =\delta^i_j,
\end{equation} 
which is extended to a map  $\Omega^p(M, \mathrm T[1]^*\mathfrak{sl}(2;\mathbb R))\times \Omega^q(M, \mathrm T[1]^*\mathfrak{sl}(2;\mathbb R))\to \Omega^p(M)$ in the obvious manner, for example
\begin{equation}
 \langle \alpha^i t_i, \beta_j \tilde t^j \rangle =\alpha^i \wedge \beta_i.
\end{equation} 

 The corresponding AKSZ theory is then nothing but the two-dimensional \(BF\) model,
\begin{equation}\label{eq:JT_action}
     S_{\text{AKSZ}}[A, \phi] = \int F^i\wedge\phi_i,
\end{equation}
where \(\phi\) is an \(\mathfrak{sl}(2;\mathbb R)^*\)-valued scalar field of mass dimension zero and \(F= \mathrm dA+\frac12[A\wedge A]\), where $A=A^i t_i$ is an \(\mathfrak{sl}(2;\mathbb R)\)-valued connection of mass dimension one.

Let us now  decompose $A^i$ under 
\begin{equation}
\begin{array}{cccccccccccc}
\mathfrak{sl}(2;\mathbb R)& \cong& \mathfrak{so}(1,1)& + &\mathbb R &+ &\mathbb R\\
\mathbf{3} & \cong&\mathbf{1}_0& + &\mathbf{1}_{+1} &+ &\mathbf{1}_{-1}\\
\end{array}
\end{equation} into $(A^a, A^2)$, where $a=0,1$. Take for Hodge structure the projector  
\begin{equation}\label{eq:2dM}
\begin{array}{cccccccccccc}
M_{\text{nls}}:& \mathfrak{sl}(2;\mathbb R)\ltimes\mathfrak{sl}(2;\mathbb R)^*[-1] &\to &\mathfrak{m}^*[-1]\\
&(t_i, \tilde t^j) &\mapsto &(Mt_i, M\tilde t^j)&=  (\mu \delta_{i}^{a}\eta_{ab}\tilde t^b, 0)
\end{array}
\end{equation}
where $\mathfrak{m}\cong \mathbb R + \mathbb R$ is the orthogonal complement of $\mathfrak{so}(1,1)\subset \mathfrak{sl}(2; \mathbb{R})$,  $\mu$ is a dimensionless parameter and  \(\eta_{ab}=\operatorname{diag}(-1,1)\) is the Minkowski metric.\footnote{We leave the degree-shift in \eqref{eq:2dM} implicit.} Note, the kernel of $M_{\text{nls}}$ is actually larger  than $\frac12\text{dim} \operatorname{T}^*[1]\mathfrak{sl}(2;\mathbb R))$, so this is a generalisation of the standard notion of a Manin pair, as described in \cite{Dimi} for general non-linear sigma models.

Further introducing a background metric $\hat g$ on $M$ and the corresponding Hodge duality operator $\hat\star$, we can then form the  Manin mass-like  term\footnote{If we were to have \(\mathfrak g=\mathfrak{so}(1,1)\) rather than \(\mathfrak{sl}(2;\mathbb R)\), we could also have a topological Manin term as in \cref{sec:4d}, corresponding to a bare cosmological constant term \(\lambda e^a\wedge e^b\epsilon_{ab}\) in the gravitational interpretation. With only the topological deformation, diffeomorphisms are preserved as in the  four-dimensional case,  \cref{sec:4d}.}  
\begin{equation}
\tfrac12\langle A^it_i + \phi_i \tilde t^i, \hat\star M_{\text{nls}}(A^it_i + \phi_i \tilde t^i)\rangle= \tfrac12\mu\eta_{ab}A^a\wedge \hat \star A^b.
\end{equation}
Adding  to the action \eqref{eq:JT_action}, this  yields the corresponding AKSZ--Manin theory:
\begin{equation}\label{eq:JT_metric_only_mass}
    S^{M_\text{nls}}_{\text{AKSZ}}[A, \phi] = \int F^i\wedge\phi_i+\tfrac12\mu\eta_{ab}A^a\wedge \hat \star A^b.
\end{equation}
 As we shall explain in the following,  \eqref{eq:JT_metric_only_mass} admits both gauge theory and gravitational interpretations.  The former is a standard non-linear sigma model with $\operatorname{AdS}_2$ target space, while the latter is Jackiw--Teitelboim gravity with a diffeomorphism-breaking background stress–energy tensor that couples to the dynamical  metric. 

\subsection{Gauge theory interpretation}
The AKSZ--Manin action \eqref{eq:JT_metric_only_mass} has the form of a Freedman--Townsend model \cite{Freedman:1980us} in which only some of the fields \(A^a\) are made auxiliary \cite{Balachandran:1981xw}. The fields \(\phi\) act as Lagrange multipliers, constraining \(F=0\). This implies that \(A\) is pure gauge,
\begin{equation}\label{eq:Apuregauge}
    A = g^{-1}d  g,
\end{equation}
for some scalar field \(g: M\to\operatorname{SL}(2;\mathbb R)\cong\operatorname{Spin}(1,2;\mathbb R) \) . Substituting \eqref{eq:Apuregauge} back into the action \eqref{eq:JT_metric_only_mass} yields
\begin{equation}
    S^{M_\text{nls}}_{\text{AKSZ}}[g] = \int \tfrac12\eta_{ab}(\Pi g^{-1}d g)^a\wedge \hat \star(\Pi g^{-1}d g)^b
\end{equation}
where
\begin{equation}
    \Pi\colon\mathfrak{so}(1,2)\to\mathfrak m
\end{equation}
is the projector to the two-dimensional subspace \(\mathfrak m\subset\mathfrak{so}(1,2)\) orthogonal to \(\mathfrak{so}(1,1)\). We conclude that the AKSZ--Manin theory, in this case,  is the non-linear sigma model on $M$ with target \(\operatorname{SO}(1,2)/\operatorname O(1,1)=\operatorname{AdS}_2\).

\subsection{Gravitational interpretation}\label{ssec:2dgrav}
As is well-known,  the $BF$ theory \eqref{eq:JT_action} is perturbatively equivalent \cite{Fukuyama:1985gg,Isler:1989hq} to Jackiw--Teitelboim gravity \cite{Teitelboim:1983ux,Jackiw:1984je} (reviewed in \cite{Mertens:2022irh,Grumiller:2002nm,Grumiller:2006rc}) under the correspondence
\begin{align}\label{eq:JT_correspondence}
    A^i_\mu&=(M_\mathrm{Pl}e^a_\mu,\tfrac12\epsilon_{ab}\omega^{ab}_\mu)&
    \phi_i&=(\lambda_a,\Phi),
\end{align}
where \(M_\mathrm{Pl}\) is an arbitrary mass scale and the Levi-Civita symbol is \(\epsilon_{01}=-\epsilon^{01}=-1\).  In terms of these variables the AKSZ action \eqref{eq:JT_action} is the Jackiw--Teitelboim gravity action
\begin{equation}
    S_{\text{JT}}[e,\omega,\Phi] = \int\epsilon_{ab}\left(\tfrac12\Phi\,R^{ab}+\tfrac12M_\mathrm{Pl}^2\Phi e^a\wedge e^b\right)+\lambda_a\wedge(\mathrm de^a+\omega^a{}_b\wedge e^b),
\end{equation}
where the two-dimensional Riemann tensor is cast as the two-form
\begin{equation}
    R^{ab}=\mathrm d\omega^{ab}
\end{equation}
and the two-dimensional dimensionless gravitational constant has been absorbed into the dilaton \(\Phi\). The fields \(\lambda\) are Lagrange multipliers that enforce the torsion-freeness of the spin connection.

Now consider the AKSZ--Manin theory, given by \eqref{eq:JT_metric_only_mass}. Under the correspondence
\begin{equation}
    \sqrt{|\hat g|}(\hat g^{-1})^{\mu\nu} = M_\mathrm{Pl}^{-2}T^{\mu\nu},
\end{equation}
the action \eqref{eq:JT_metric_only_mass} becomes
\begin{equation}
     \int\epsilon_{ab}\left(\tfrac12\Phi\,R^{ab}+\tfrac12M_\mathrm{Pl}^2\Phi e^a\wedge e^b\right)+\lambda_a\wedge(\mathrm de^a+\omega^a{}_b\wedge e^b) + \tfrac12\mu T^{\mu\nu}e^a_\mu e_{a\nu},
\end{equation}
which  amounts to adding a background stress--energy tensor for the string-frame metric (\emph{not} the Einstein-frame metric).

There are two immediate generalisations or variations of the above construction. First, JT gravity belongs to the broader class of  dilaton gravity theories in two dimensions, which may be formulated as more general Poisson sigma models \cite{Ikeda:1993fh,Schaller:1994es,Schaller:1994uj} (reviewed in \cite{Schaller:1995xk,contreras}), to which we may similarly add mass-like Manin terms \cite{Arvanitakis:2025nyy}. The analysis goes through essentially unchanged, as briefly discussed in \cref{app:PSM}. Second, for all Poisson sigma models  (including JT gravity) we may consider other Hodge structures $M$ and corresponding Manin mass-like terms.  In addition to the string-frame stress--energy tensor density,  described above, we may add a  dilaton potential, or relax the torsion-free condition enforced by the Lagrange multiplier $\lambda$. We now give examples of this kind for the case of JT gravity.

\subsection{Variant mass terms for Jackiw--Teitelboim gravity}

Note that \eqref{eq:2dM} is obviously not the unique choice of Hodge structure.  Indeed, by  consider alternative  Hodge structures $M$ we can construct other gauge and gravity models all with mother action,
\begin{equation}
S^{M}_{\text{AKSZ}}[A, \phi] = \int F_i\wedge\phi^i+\tfrac12  \langle A^it_i + \phi_i \tilde t^i, \hat\star M(A^it_i + \phi_i \tilde t^i)\rangle.
\end{equation}

An obvious example is given by  
\begin{equation}\label{eq:2dMpcm}
\begin{aligned}
M_{\text{pcm}}\colon \mathfrak{sl}(2;\mathbb R)\ltimes\mathfrak{sl}(2;\mathbb R)^*[-1] &\to \mathfrak{sl}(2;\mathbb R)^*[-1] \\
(t_i, \tilde t^j)&\mapsto  Mt_i = \mu \eta_{ij}\tilde t^j,
\end{aligned}
\end{equation}
where \(\eta^{ij}=\operatorname{diag}(-1,1,1)\) is the \(\mathfrak{sl}(2;\mathbb R)\) Killing form. This is closer in form to the original Manin theories, since $\mathfrak{sl}(2;\mathbb R)\ltimes\mathfrak{sl}(2;\mathbb R)^*[-1]$ and  $\mathfrak{sl}(2;\mathbb R)^*[1]$ constitute a Manin pair in the conventional sense.

This yields the AKSZ--Manin theory
\begin{equation}
\begin{split}
    S^{M_\text{pcm}}_{\text{AKSZ}}[A, \phi] &= \int F_i\wedge\phi^i+\tfrac12  \langle A^it_i + \phi_i \tilde t^i, \hat\star M_{\text{pcm}}(A^it_i + \phi_i \tilde t^i)\rangle\\
    &= \int F_i\wedge\phi^i+\tfrac12\mu\eta^{ij}A_i\wedge\star A_j.
    \end{split}
\end{equation}
Again, $A$ is pure gauge, but in this case the image on $M_\text{pcm}$ is the full $\mathfrak{sl}(2;\mathbb R)$ and so we obtain the  \(\operatorname{SL}(2;\mathbb R)\) principal chiral model in the gauge theoretic  interpretation.

Translating into the gravitational picture  we pick up a coupling of the background stress-energy density to the spin-connection:
\begin{multline}
    S^{M_\text{pcm}}_{\text{AKSZ}}[e, \omega, \Phi]  =\int\epsilon_{ab}\left(\tfrac12\Phi\,R^{ab}+\tfrac12M_\mathrm{Pl}^2\Phi e^a\wedge e^b\right)+\lambda_a\wedge(\mathrm de^a+\omega^a{}_b\wedge e^b) \\
    + \tfrac12T^{\mu\nu}\eta_\mu^a \eta_{b\nu}+M_\mathrm{Pl}^{-2}T^{\mu\nu}\omega^{ab}_\mu\omega_{ab\nu}.
\end{multline}
 
 The larger kernel of $M_{\text{pcm}}$ yields a nonminimal coupling to the torsionful spin connection \(\omega^{ab}_\mu\), resulting in a form of Einstein--Cartan gravity with background stress-tensor density.

Alternatively, we can take 
\begin{equation}\label{eq:2dMalt}
\begin{aligned}
M_{\text{twt}}\colon \mathfrak{sl}(2;\mathbb R)\ltimes\mathfrak{sl}(2;\mathbb R)^*[-1] &\to \mathfrak{sl}(2;\mathbb R)\\
(t_i, \tilde t^j)&\mapsto  M\tilde t^j = \mu \eta^{ij} t_j,
\end{aligned}
\end{equation}
which yields a two-dimensional third-way theory \cite{Arvanitakis:2015oga, Deger:2021ojb}, 
\begin{equation}
    S_{\text{AKSZ}}^{M_{\text{twt}}}[A, \phi] = \int F_i\wedge\phi^i+\tfrac12\mu\eta^{ij}\phi_i\wedge\star\phi_j.
\end{equation}

On the gravitational side, this  introduces  to the Jackiw--Teitelboim action \eqref{eq:JT_action} a mass-like term for the would-be Lagrange multiplier \(\lambda^a\), 
\begin{equation}
    \int\epsilon_{ab}\left(\tfrac12\,\Phi R^{ab}+\tfrac12M_\mathrm{Pl}^2\Phi e^a\wedge e^b+\tfrac12T\Phi^2\right)+\lambda_a\wedge(\mathrm de^a+\omega^a{}_b\wedge e^b) + \tfrac12T\lambda_a\lambda^a,
\end{equation}
where, in addition to \eqref{eq:JT_correspondence}, we define the two-form
\begin{equation}
    T= \mu\operatorname{vol}_{\hat g}.
\end{equation}
This appears as a quadratic term in the dilaton potential together with a relaxation of the torsion-free condition (to a Gaussian smearing thereof) that was formerly enforced by the Lagrange multipliers \(\lambda_i\).

\section{Three dimensions}
The  well-known Chern--Simons descriptions  of  gravity in three dimensions have gauge groups \(\mathfrak{so}(3;\mathbb C)=\mathfrak{sl}(2;\mathbb C)\) (for dS), \(\mathfrak{so}(2,2)=\mathfrak{sl}(2;\mathbb R)\oplus\mathfrak{sl}(2;\mathbb R)\) (for AdS), and \(\mathfrak{iso}(1,2;\mathbb R) = \mathrm T^*\mathfrak{sl}(2;\mathbb R)\) (for Poincar\'e). 
If one does not break three-dimensional Lorentz symmetry \(\mathfrak{so}(1,2)\), then the Manin pairs are unique except for the last case, for which there are two possibilities:
 \begin{equation}
 (\mathrm T^*\mathfrak{sl}(2;\mathbb R),\mathfrak{sl}(2;\mathbb R)),\qquad (\mathrm T^*\mathfrak{sl}(2;\mathbb R),\mathfrak{sl}(2;\mathbb R)^*).
 \end{equation}

The structure constants are given by 
\begin{align}
    \tilde h^{abc}&= 0 &
    M^{ab}&=\mu\eta^{ab} &
    f_{abc} &= \epsilon_{abc} &
    \tilde f_{abc}&=\nu\epsilon_{abc}
\end{align}
with \(\nu\in\{-1,0,+1\}\), where now the indices $a,b,c\in \{0,1,2\}$ are the dreibein indices in the adjoint of $\mathfrak{sl}(2; \mathbb R)\cong \mathfrak{so}(1,2)$. (We freely raise and lower the dreibein indices using the Minkowski metric $\eta^{ab}$ and so do not notationally distinguish between $\mathfrak{sl}(2; \mathbb R)$ and $\mathfrak{sl}(2; \mathbb R)^*$ indices.)

The first three  cases have been treated in \cite{Borsten:2024pfz}, so we just briefly summarise them here. Making the identifications  
\begin{align}
	e^a &= kM_\mathrm{Pl}^{-1}\tilde A^a &
	\omega^{ab} &= -\epsilon^{abc}A_c \\
	\Lambda &= -\frac{\nu M_\mathrm{Pl}^2}{4k^2}&
	T^{\mu\nu} &= \frac{\mu M_\mathrm{Pl}^2}{k^2}(\hat g^{-1})^{\mu\nu}\sqrt{|\det\hat g|},
\end{align}
the Chern--Simons Manin theory action \eqref{eq:Chern--Simons-manin-action} takes the gravitational form
\begin{equation}
	S[e,\omega] = \tfrac12M_\mathrm{Pl} \int \epsilon_{abc}\left(e^a\wedge R^{bc}-\tfrac13\Lambda e^a\wedge e^b\wedge e^c\right)+\tfrac12\int\mathrm d^3x\, T^{\sigma\tau}\eta_{ab}e^a_\sigma e^b_\tau.
\end{equation}
The first two terms are just  3d Einstein--Cartan gravity. The final terms follows from the Manin mass-like deformation   and corresponds to a background stress--energy density minimally coupled to the dynamical metric.

Here we discuss the fourth case with Manin pair $(\mathrm T^*\mathfrak{sl}(2;\mathbb R),\mathfrak{sl}(2;\mathbb R)^*)$. Just as for $(\mathrm T^*\mathfrak{sl}(2;\mathbb R),\mathfrak{sl}(2;\mathbb R))$, the cosmological constant is vanishing since $v=0$. 

Thus, under the correspondence
\begin{align}
    e^a &= kM_\mathrm{Pl}^{-1}\tilde A^a &
    \omega^{ab} &= -\epsilon^{abc}A_c &
    S^{\mu\nu} &= k^{-2}\mu\sqrt{|\det\hat g|}(\hat g^{-1})^{\mu\nu},
\end{align}
we again arrive at  Poincar\'e  Einstein-Hilbert gravity with a background density $S^{\mu\nu}$, 
\begin{equation}\label{eq:3d_Einstein_Cartan}
    S[e,\omega]=\tfrac12M_\mathrm{pl}\int\epsilon_{abc}e^a\wedge R^{bc}-\tfrac14\int\mathrm d^3x\,S^{\mu\nu}\omega_{ab\mu}\omega^{ab}_\nu.
\end{equation}
However, we see the backgorund density has a nonminimal coupling to the spin connection. Consequently, it is no longer torsion-free
\begin{equation}
d e + \omega \wedge e  \propto \omega^\mu S_{\mu\nu} dx^\nu. 
\end{equation}
Thus, this is a form of Einstein--Cartan theory with spin-coupling.

\section{Four dimensions}\label{sec:4d}

Here we consider the ASKZ theory with four-dimensional (pseudo-)Riemannian  source manifold $\Sigma$ and graded Lie algebra
\begin{equation}\label{eq:MMgLA}
    \mathfrak d = \mathrm T^*[1]\mathfrak{so}(2,3) = (\mathfrak{so}(2,3)^*[1] \to \mathfrak{so}(2,3)).
\end{equation}
In this section, \(i,j,\dotsc\in\{-1,0,1,2,3\}\) range over the defining vector representation $V\cong\mathbb R^{2,3}$ of the (anti-)de~Sitter algebra \(\mathfrak{so}(2,3)\), whose signature we take to be \(--+++\). The indices \(a,b\in\{0,1,2,3\}\) range over vierbein indices.

Let \(\mathtt b^{ij},\mathtt a_{ij}\) be a homogeneous basis  for \eqref{eq:MMgLA}, with the nontrivial brackets  \([\mathtt a,\mathtt a]\) and \([\mathtt a,\mathtt b]\)  given by the adjoint and coadjoint \(\mathfrak{so}(2,3)\)-representations, respectively; the Lie bracket \([\mathtt b,\mathtt b]\) vanishes. On $\mathfrak d$, we have an invariant inner product of degree \(1\) given with respect to the above basis by  
\begin{equation}
    \langle \mathtt a_{ij}, \mathtt b^{kl} \rangle =\delta_i^{[k} \delta_j^{l]},
\end{equation} 
which  extends to an inner product on   $\Omega^p(M, \mathfrak d)$ in the obvious manner, for example 
\begin{equation}
 \langle \alpha^{ij} \mathtt a_{ij} , B_{kl} \mathtt b^{kl}\rangle = \int \alpha^{ij} \wedge \beta_{ij},
\end{equation} 
where $\alpha, \beta$ are 2-forms. 

The corresponding AKSZ action is then the \(BF\) action
\begin{equation}\label{eq:ASKZ_MM}
    S_{\text{AKSZ}}[A,B] = \int B_{ij}\wedge F^{ij},
\end{equation}
where \(B,A\) are \(\mathfrak{so}(2,3)^*\)-, \(\mathfrak{so}(2,3)\)-valued 2-, 1-forms of mass dimensions 2 and 1, respectively, and
\begin{equation}
    F^{ij}\coloneqq \mathrm dA^{ij}+\frac12[A\wedge A]^{ij}
\end{equation}
is the \(\mathfrak{so}(2,3)\)-valued field strength.

 To define a (degenerate\footnote{In the sense that the dimension of \(\mathfrak g\) is less than half the dimension of \(\mathfrak d\), as in the non-linear sigma model discussed in \cref{sec:2d}.}) Hodge structure, we pick a preferred timelike direction in \(\mathfrak{so}(2,3)\), breaking the symmetry from the AdS group to the Lorentz group \(\mathfrak{so}(1,3)\).   We take the \(\mathfrak{so}(2,3)\)-breaking vector to  be \(\delta^i_{-1}\) for definiteness and  let \(\mathfrak g=\mathfrak{so}(1,3)\subset\mathfrak d\) denote the corresponding preserved subalgebra.

Generically, choosing a \(\mathfrak g\)-equivariant Hodge structure 
\begin{equation}
\begin{aligned}
M\colon\mathfrak d  \to \mathfrak d 
\end{aligned}
\end{equation}
and a background metric $\hat g$ on $\Sigma$, we can add a masslike term to obtain the corresponding ASKZ--Manin theory
\begin{equation}
S^{M}_{\text{AKSZ}}[A,B] =S_{\text{AKSZ}}[A,B] -\frac12 \int \langle \Phi, \hat \star M \Phi\rangle,
\end{equation}
where we have introduced the inhomogenous $\mathfrak d$-valued polyform superfield $\Phi = B_{ij} \mathtt b^{ij}+ A^{ij} \mathtt a_{ij}$.

However, since $B$ is a $\frac12d$-form in this case, if we choose $M$ such that $ \operatorname{im}~ M \subseteq \mathfrak{so}(2,3) \subseteq \ker M$, then we also have a topological Manin deformation
\begin{equation}
\begin{split}
S^{M~\text{top}}_{\text{AKSZ}}[A,B] &=S_{\text{AKSZ}}[A,B] -\frac12 \int \langle \Phi,  M \Phi\rangle,\\
&=S_{\text{AKSZ}}[A,B] -\frac12 \int  B_{ij}\wedge M^{ij, kl} B_{kl}. 
\end{split}
\end{equation}
This will provide a direct connection to Einstein gravity with a cosmological constant.

In particular, let us define
\begin{equation}\label{eq:4dMA}
\begin{aligned}
M\colon \mathfrak d &\to \mathfrak{so}(1,3)\\
(\mathtt b^{ij},\mathtt a_{kl})&\mapsto  M \mathtt b^{ij}=  \tfrac12\alpha\epsilon^{-1ijkl}\mathtt a_{kl},
\end{aligned}
\end{equation}
where \(\alpha\) is an arbitrary parameter.

In this case, the AKSZ--Manin theory with topological deformation is  
\begin{equation}\label{eq:4dMAaction}
    S^{M~\text{top}}_{\text{AKSZ}}[\Phi] =\int B_{ij}\wedge F^{ij}-\frac14\alpha  \epsilon^{-1ijkl}B_{ij}\wedge  B_{kl}.
\end{equation}
This is precisely the $BF$ formulation of MacDowell--Mansouri gravity \cite{MacDowell:1977jt,Stelle:1979va} gravity given in \cite{Freidel:2005ak}, where it is shown to be    equivalent to Einstein-Hilbert gravity with negative cosmological constant (up to a topological term given by the integrated Euler characteristic).
Indeed, upon rewriting the connection \(A^{ij}\) in terms of the vierbein \(e^a\) and spin connection \(\omega^{ab}\),
\begin{align}
	A^{-1a} &= \mu e^a,
	\\
	A^{ab} &= \omega^{ab},
\end{align}
 the field strength \(F^{ij}\) decomposes into the torsion tensor \(T^a\) and Riemann curvature tensor \(R^{ab}\),
\begin{align}
\begin{aligned}
	F^{-1a} &= \mu T^a\,,
	&
	T^a &= de^a + \omega^a{_b}\wedge e^b,
	\\
	F^{ab} &= R^{ab} - \mu^2 e^a\wedge e^b\,,
	&
	R^{ab} &= d\omega^{ab} + \omega^{ac}\wedge\omega_c{^b},
\end{aligned}
\end{align}
where   \(\mu\) is mass-dimension-\(1\) constant. Writing the action \eqref{eq:4dMAaction} in terms of $e^a, \omega^{ab}, B_{ij}$  we  obtain
\begin{equation}\
	S^{M~\text{top}}_{\text{AKSZ}}[e,\omega,B] 
	=
	\int 2\mu B_{-1a}\wedge T^a 
	+ B_{ab}\wedge\big(R^{ab} - \mu^2 e^a\wedge e^b\big)
	- \frac{\alpha}{4}\epsilon^{abcd}B_{ab}\wedge  B_{cd}\,.
\end{equation}
Here, \(B_{-1a}\) serves as a Lagrange multiplier killing the torsion,  $T^a=0$, which fixes $\omega^{ab}$ to be the spin-connection (assuming $e^a$ is invertible). Similarly,  \(B_{ab}\)  is auxiliary with algebraic equation of motion
\begin{equation}\label{Baux}
\big(R^{ab} - \mu^2 e^a\wedge e^b\big)
	- \frac{\alpha}{2}\epsilon^{abcd}  B_{cd}=0. 
\end{equation}
 
 Upon integrate out the Lagrange multiplier \(B_{-1a}\) and eliminating \(B_{ab}\) via \eqref{Baux} we obtain,
\begin{equation}
	S^{M~\text{top}}_{\text{AKSZ}}[e,\omega] 
	=
	 - \frac{1}{4\alpha} \int \epsilon_{abcd}\big(R^{ab} - \mu^2e^a\wedge e^b\big)\wedge\big(R^{cd} - \mu^2e^c\wedge e^d\big)\,.
\end{equation}
Let us further define the gravitational constant \(G\) and cosmological constant \(\Lambda\) in terms of \(\alpha\) and \(\mu\) as
\begin{align}
	G &\coloneq \alpha\mu^{-2}\,,
	&
	\Lambda &\coloneq 3\mu^2\,.
\end{align}
so that
\begin{multline}
	S^{M~\text{top}}_{\text{AKSZ}}[e,\omega(e)] 
	=
	\frac{1}{2G}\int \epsilon_{abcd}e^a\wedge e^b\wedge R^{cd} - \frac{\Lambda}{3!}\epsilon_{abcd}e^a\wedge e^b\wedge e^c\wedge e^d 
	\\
	- \frac{3}{2\Lambda}\epsilon_{abcd}R^{ab}\wedge R^{cd}\,.
\end{multline}
The first two terms correspond to Einstein gravity with a negative cosmological constant \(\Lambda\), whereas the final term is  topological  and does not contribute to the equarions of motions.

We conclude that the purely topological four-dimensional AKSZ theory  with gauge algebra $\mathrm{T}^*[1]\mathfrak{so}(2,3)$ may be deformed with a Manin term to produce dynamical Einstein gravity with cosmological constant. The Euclidean and (Euclidean/Lorentzian) dS cases follow precisely the same logic. We expect the supergravity generalisation for  Lorentzian AdS to follow straightforwardly, but we leave this for future work.

The background dependent Manin deformation similarly yields 
\begin{equation}
    S^{M}_{\text{AKSZ}}[\Phi] =\int B_{ij}\wedge F^{ij} -\frac14\alpha  \epsilon^{-1ijkl}B_{ij}\wedge \hat \star B_{kl}.
\end{equation}
%
The action \eqref{eq:4dMAaction} is remains  similar to the \(BF\) formulation of MacDowell--Mansouri gravity, except that the \(B^2\) term contains a Hodge star with respect to the background.

Much of the physics remains unchanged, however. Following precisely the same logic, we arrive at
\begin{equation}\label{eq:4dMAaction_red}
    S^{M}_{\text{AKSZ}}[\Phi] =\int B_{ab}\wedge F^{ab} -\frac14\alpha  \epsilon^{abcd}B_{ab}\wedge \hat \star B_{cd}.
\end{equation}
 Eliminating \(B^{ab}\) from the above action, we find
\begin{equation}
\begin{split}
    S^{M}_{\text{AKSZ}}[\Phi]  &= \frac1{4\alpha}\int F^{ab}\wedge\hat\star F^{ab} \\
    &= -\frac1{2G}\int\left(R^{ab}\wedge\hat\star(e^c\wedge e^d)
    -\frac16\Lambda e^a \wedge e^b \wedge\hat \star(e^c\wedge e^d)
    \right)\epsilon_{abcd}\\&+\frac3{4G\Lambda}\int R^{ab}\wedge\hat \star R^{cd}\epsilon_{abcd},
\end{split}
\end{equation}
where we have defined
\begin{align}
    G&\coloneqq \alpha\mu^{-2} &
    \Lambda &\coloneqq 3\mu^2.
\end{align}
The first term is a would-be Palatini action for  Einstein--Hilbert gravity, except that the Ricci `scalar' is built involving both the dynamical metric \(g_{\mu\nu}=e^a_\mu e_{a\nu}\) as well as the background metric \(\hat g_{\mu\nu}\). The second term is a version of background stress--energy tensor, and the third curvature-squared term would be the first Pontrjagin class if the background and dynamical metric were identified.

\section*{Acknowledgements}
H.K. was supported by the Leverhulme Research Project Grant \textsc{rpg}--2021--092.
The authors thank    Alexandros Spyridion Arvanitakis\orcidlink{0000-0002-7844-5574} and Maxim Grigoriev\orcidlink{0000-0002-2368-7445} for helpful discussions.

\appendix 

\section{Manin Poisson sigma models}\label{app:PSM}

A Poisson manifold is a smooth manifold equipped with a (2,0)-tensor field \(\Pi\) that defines a Poisson bracket on its algebra of functions (see~\cite{MR1636305,MR2906391,MR4328925} for details). Consider the Poisson manifold \(X=(\mathbb R^3,\Pi)\) coordinatised by \(\{y^0,y^1,y^2\}\) with
\begin{align}
    \Pi^{01}&=\frac12V(y^2)&
    \Pi^{02}&=y^1&
    \Pi^{12}&=-y^0.
\end{align}
Then its (degree-shifted) cotangent bundle \(\mathrm T[1]^*X\) admits the structure of a Lie algebroid. The corresponding AKSZ theory is the Poisson sigma model
\begin{equation}
    S = \int A\wedge\mathrm d\phi^i+\tfrac12\Pi^{ij}(\phi)A_i\wedge A_j.
\end{equation}
Using
\begin{align}
    A^i_\mu&=(M_\mathrm{Pl}e^a_\mu,\tfrac12\epsilon_{ab}\omega^{ab}_\mu)&
    \phi_i&=(\lambda_a,\Phi),
\end{align}
the action takes the gravitational form
\begin{multline}
    S[e,\omega,\Phi,\lambda] = \int\epsilon_{ab}\left(\tfrac12\Phi\,R^{ab}+\tfrac12M_\mathrm{Pl}^2V(\Phi)e^a\wedge e^b\right)+\lambda_a\wedge(\mathrm de^a+\omega^a{}_b\wedge e^b).
\end{multline}

To the Poisson sigma model we can add the mass term
\begin{equation}
    S \ni \int \frac12\mu\eta^{ij}A_i\wedge\star A_j.
\end{equation}
This produces a propagating sigma model. Gravitationally, this corresponds to adding a coupling
\begin{equation}
    S \ni \int\mathrm d^2x\, (T^{\mu\nu}g_{\mu\nu} + M_\mathrm{Pl}^{-2}T^{\mu\nu}\omega^{ab}_\mu\omega_{ab\nu})
\end{equation}
where
\begin{equation}
    T^{\mu\nu}=\sqrt{|\det\hat g|}(\hat g^{-1})^{\mu\nu}.
\end{equation}
So this amounts to adding a background stress--energy tensor together with a nonstandard coupling to the spin connection.

\bibliographystyle{unsrturl}
\bibliography{biblio}

\end{document}